\documentclass[aps,prl,twocolumn,10pt,groupedaddress,showpacs]{revtex4-1}

\usepackage{graphicx}
\usepackage{amssymb}
\usepackage{amsthm}
\usepackage{amsfonts,amsmath,latexsym,bbm}
\usepackage{ae,aecompl}                          
\usepackage{dsfont}
\usepackage{tensor}
\usepackage{color}
\usepackage{tikz-cd}
\usepackage{tikz}
\usetikzlibrary{decorations.markings}
\usetikzlibrary{matrix,arrows}
\usepackage{natbib}

\newcommand{\de}{\mathrm d}
\newcommand{\inv}{^{-1}}
\newcommand{\tsr}[1]{\overset\leftrightarrow{#1}}
\newcommand{\ext}{_{\rm ext}}
\newcommand{\tot}{_{\rm tot}}

\renewcommand{\vec}[1]{\boldsymbol{#1}}
\newcommand{\h}{\hspace{1pt}}
\newcommand{\lar}[1]{\textnormal{\mbox{\large $#1$}}}

\definecolor{darkred}{rgb}{0.8,0,0}

\begin{document}


\title{Relativistic covariance of Ohm's law}


\author{R.~Starke}
\affiliation{Department of Computational Materials Physics, University of Vienna, \\ Sensengasse 8/12, 1090 Vienna, Austria}
\affiliation{Institute for Theoretical Physics, TU Bergakademie Freiberg, Leipziger Stra\ss e 23, 09596 Freiberg, Germany}

\author{G.\,A.\,H.~Schober}
\affiliation{Institute for Theoretical Physics, Heidelberg University, Philosophenweg 19, 69120 Heidelberg, Germany}


\date{\today}

\begin{abstract}
The derivation of Lorentz-covariant generalizations of Ohm's law has been a long-term issue in theoretical physics with deep implications for the study of relativistic effects in optical and atomic physics. In this article, we propose an alternative route to this problem, which is motivated by the tremendous progress in first-principles materials physics in general and ab initio electronic structure theory in particular. We start from the most general, Lorentz-covariant first-order response law, which is written in terms of the fundamental response tensor~$\chi\indices{^\mu_\nu}$ relating induced four-currents to external four-potentials. By showing the equivalence of this description to Ohm's law, we prove the validity of Ohm's law in every inertial frame. We further use the universal relation between $\chi\indices{^\mu_\nu}$ and the microscopic conductivity tensor $\sigma_{k\ell}$ to derive a fully relativistic transformation law for the latter, which includes all effects of anisotropy and relativistic retardation. In 
the special case of a constant, scalar conductivity, this transformation law can be used to rederive a standard textbook generalization of Ohm's law.
\end{abstract}

\pacs{03.30.+p, 03.50.De}

\maketitle

{\itshape Introduction.}---Formulated as a simple, yet ingeneous equality, capable of explaining a plethora of experimental data already at the time of its discovery, 
Ohm's law \cite{Ohm1827} soon found its technological application in the engineering of nineteenth century telegraph systems \cite{Morse, Butrica}. 
Since then it has been used in nearly every branch of physical sciences to describe such different systems as neuron 
cells in medical physiology \cite{Hodgkin, Lytton}, black hole membranes in astrophysics \cite{Parikh, Thorne}, and recently even strongly interacting $(2 + 1)$-dimensional conformal field theories with anti-de Sitter space duals \cite{String}.
Although it had been believed for a long time that Ohm's law would break down at the atomic scale, it was demonstrated in 2012 to hold in silicon
wires only four atoms wide \cite{Weber}, raising the prospect of further applications in atomic-scale logic circuits \cite{Ryu, Ferry, Fuechsle}. 
On the theoretical side, the problem of deriving Ohm's law from microscopic models (as formulated e.g.~by Peierls \cite{Peierls}) 
is attracting continuous interest \cite{Chernov}. In plasma physics, a generalized Ohm law is 
used to describe an electrically conducting, moving medium in the presence of magnetic fields \cite{Krall, Lorrain}. 
Indeed, in the magnetohydrodynamic description, where Maxwell's equations govern the electromagnetic fields while the fluid is subject to
energy and momentum conservation, Ohm's law expresses the coupling between the electromagnetic fields and the fluid variables \cite{Lichnerowicz, Palenzuela}. 
This  becomes especially relevant in astrophysics and cosmology, where relativistic plasmas are used to describe 
the formation of black holes, the generation of jets and gravitational waves \cite{Meier} as well as the evolution of the early universe \cite{Holcomb}. 
Consequently, an intense research activity had been focused on the derivation of relativistic generalizations of Ohm's law 
\cite{Ardavan, Bekenstein, Blackman, Gedalin, Khanna, Kremer, Kandus, Sherlock, Ahmedov2011, Bucciantini}, a problem 
which has gradually reached the modern textbook literature (see e.g.~\cite[Section~13.14]{Tsamparlis}).

In a broader sense, relativistic approaches to condensed matter physics are also highly important for practical reasons, 
since they allow for the investigation of moving media \cite{Leonhardt, Mackay07}. The latter also play a significant r\^{o}le in classical optics, where e.g.~the refractive index of a moving medium is determined by the Fresnel drag coefficient \cite{Joos, Bergmann, Laue}. In this context, the relativistic transformation properties of the microscopic conductivity tensor are particularly important. 
In terms of the conductivity tensor, electrodynamics of materials can be formulated as a single-susceptibility theory \cite{Cho10,Melrose}, 
i.e., all electromagnetic response functions can be expressed analytically in terms of the conductivity tensor \cite[Sec.~6.5]{ED1}. Hence, the 
transformation properties of the conductivity tensor essentially determine all other transformation properties of electromagnetic material responses. 
Finally, in the ab initio materials physics and electronic structure physics communities, a relativistic formulation of Ohm's law 
is becoming increasingly relevant, since it turned out that the corresponding optical response function is also influenced by 
relativistic effects \cite{Liberman, Neutze, Boeij, Romaniello, Lee, Dani}.

The present article is concerned with a more restricted problem, the generalization of Ohm's law to {\itshape special} relativistic covariance. On a macroscopic scale, Ohm's law relates the induced electric current density $\vec j_{\rm ind} \equiv \vec j$ 
through the direct conductivity $\sigma$ to an externally applied electric field by
\begin{equation} \label{eq_naive_Ohm}
\vec j = \sigma \vec E\ext \,,
\end{equation}
or through the proper conductivity $\widetilde\sigma$ to the total electric field by \vspace{-5pt}
\begin{equation} \label{eq_naive_Ohm1}
\vec j = \widetilde\sigma \vec E\tot \,.
\end{equation}
In the following, this difference does not play any r\^{o}le, because the transformation properties of the direct and proper conductivities coincide.
Microscopically, however, Ohm's law has to be interpreted as a non-local convolution (see e.g.~\cite[Eq.~(6.29)]{Bruus}, \cite[Eqs.~(3.167) and (3.185)]{Giuliani})
\begin{equation} \label{eq_ohm}
j_k(x) = \int \! \de^4 x' \, \sigma_{k\ell}(x, x') E_\ell(x') \,,
\end{equation}
where $x \equiv x^\mu = (ct, \vec x)$ and $\de^4 x = \de x^0 \h \de^3 \vec x$. We choose the Minkowski metric as
$
 \eta_{\mu\nu} = \mathrm{diag}(-1, 1, 1, 1),
$
such that all spatial indices can be written as lower-case indices, and we sum over all doubly appearing indices. Throughout this article, we will always refer to Eq.~\eqref{eq_ohm} as Ohm's law.
From the relativistic point of view, the problem with Ohm's law apparently is that it relates the spatial part $\vec j$
of the four-vector $j^\mu=(c\rho,\vec j)$ to the spatial three vector $E_i=cF^{0i}$, 
which is part of the second-rank field strength tensor $F^{\mu\nu}=\partial^\mu A^\nu-\partial^\nu A^\mu$.
Hence, it is not obvious how Eq.~\eqref{eq_naive_Ohm} squares with the usual relativistic transformation laws. We will show below that Ohm's law in the form \eqref{eq_ohm} is nevertheless relativistically covariant, meaning that it has the same form in every inertial frame. The reason for this unexpected covariance of a seemingly non-relativistic response law is that the conductivity tensor itself obeys a complicated (non-tensorial) transformation law, which we will derive explicitly below.

In order to achieve this goal, we will not try to interpret Ohm's law as the spatial part of some covariant, four-dimensional Ohm's law, as is done
e.g.~in \cite[Sec.~1.5.6]{Melrose1Book}. Instead, we start from the following linear relation, 
which is well known in microscopic condensed matter physics, ab initio materials physics and plasma physics 
(see e.g.~\cite[Chap.~1]{Melrose1Book}, \cite[Sec.~5]{ED1}, \cite[Chap.~8]{Altland} or \cite[Sec.~I]{Adler}):
\begin{equation}\label{eq_genResplaw}
j^\mu(x)=\int\!\de^4 x'\,\chi^{\mu}_{~\nu}(x,x') A^\nu(x') \,.
\end{equation}
Here, $\chi^{\mu}_{~\nu}$ denotes the fundamental response tensor, which relates the induced four-current $j^\mu$ to the applied four-potential $A^\nu = (\varphi/c,\vec A)$. In fact, since the four-potential contains the complete information
about the externally applied fields, Eq.~\eqref{eq_genResplaw} constitutes the most general first-order response
relation, which incorporates all effects of inhomogeneity, anisotropy and relativistic retardation \cite{ED1}. 
The relation \eqref{eq_genResplaw} is relativistically covariant per constructionem, because it relates the relativistic four-vectors $j^\mu$
and $A^\nu$. In the following, we will {\itshape derive} Ohm's law \eqref{eq_ohm} from Eq.~\eqref{eq_genResplaw} in a relativistic setting. Based on the tensorial transformation law for the Lorentz tensor $\chi\indices{^\mu_\nu}$ and  the universal relation between $\chi\indices{^\mu_\nu}$ and $\sigma_{k\ell}$, we will further deduce the most general relativistic transformation law for the conductivity tensor. In the special case of a constant, scalar conductivity, we will show that the resulting equation can be used to rederive a standard textbook generalization of Ohm's law. Finally, we will discuss some problems of the standard interpretation of the relativistic generalization of Ohm's law, which are resolved by our approach.

\begin{figure}[t]
\begin{center}
\includegraphics{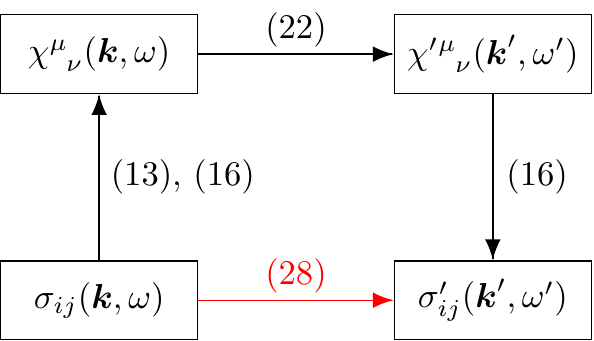}
\end{center}
\caption{Universal relations and transformation laws. The arrow labels refer to the equation numbers in the text. \label{fig_square}}
\end{figure}

{\itshape Relativistic derivation of Ohm's law.}---As mentioned in the introduction, all (linear) electromagnetic response properties can be derived from the fundamental response tensor, which can be defined as the functional derivative of the induced four-current with respect to the external four-potential \cite{Altland, Melrose1Book, ED1},
\begin{equation}\label{eq_fundconstr}
\chi^{\mu}_{~\nu}(x,x') = \frac{\delta j^\mu(x)}{\delta A^\nu(x')} \,.
\end{equation}
The current $j^\mu$ has to be invariant under gauge transformations, $A^\mu \mapsto A^\mu + \partial^\mu f$, and has to fulfill 
the continuity equation, $\partial_\mu j^\mu=0$. This implies that the fundamental response functions obey the constraints \cite{Altland,Melrose1Book}:
\begin{align}
\partial_\mu\chi\indices{^\mu_\nu}(x,x')&=0\,,  & \partial'^\nu\chi\indices{^\mu_\nu}(x,x')&=0 \label{con2}\,.
\end{align}
In the following, we assume homogeneity in space and time, such that
\begin{equation}
 \chi\indices{^\mu_\nu}(x, x') = \chi\indices{^\mu_\nu}(x - x') \,,
\end{equation}
or in the Fourier domain,
\begin{equation}
 \chi\indices{^\mu_\nu}(k, k') = \chi\indices{^\mu_\nu}(k) \h \delta^4(k - k')
\end{equation}
with $k = (\omega/c, \vec k)$. In particular, Eq.~\eqref{eq_genResplaw} then simplifies in the Fourier domain as
\begin{equation}
 j^\mu(k)=\chi\indices{^\mu_\nu}(k) \h A^\nu(k) \,.
\end{equation}
In the position/frequency domain, we can write the constraints \eqref{con2} equivalently as
\begin{align}
\chi^0_{~\ell}(\vec x - \vec x'; \omega) &= \frac{c}{\mathrm i \omega} \h \frac{\partial}{\partial x_k} \, \chi_{k\ell}(\vec x - \vec x'; \omega) \,, \label{eq_constr1}\\[3pt]
\chi_{k0}(\vec x - \vec x'; \omega) &= \frac{c}{\mathrm i \omega} \h \frac{\partial}{\partial x'_\ell} \, \chi_{k\ell}(\vec x - \vec x'; \omega) \,, \label{eq_constr2} \\[3pt]
\chi\indices{^0_0}(\vec x - \vec x'; \omega) &= -\frac{c^2}{\omega^2} \h \frac{\partial}{\partial x_k} \frac{\partial}{\partial x'_\ell} \, \chi_{k\ell}(\vec x - \vec x'; \omega) \,. \label{eq_constr3}
\end{align}
In the momentum/frequency domain, the fundamental response tensor can therefore be written explicitly as \cite[Chap.~1]{Melrose1Book} \vspace{-5pt}
\begin{equation}\label{generalform}
\chi^\mu_{~\nu}(\vec k, \omega)=
\left( \!
\begin{array}{rr} -\lar{\frac{c^2}{\omega^2}} \, \vec k^{\rm T} \, \tsr \chi \, \vec k & \lar{\frac{c}{\omega}} \, \vec k^{\rm T} \, \tsr \chi \, \\[10pt] -\lar{{\frac{c}{\omega}}} \, \tsr \chi \, \vec k & \, \tsr \chi \, 
\end{array} \right).
\end{equation}
Thus, the constraints \eqref{con2} allow for the reconstruction of the complete fundamental response tensor $\chi\indices{^\mu_\nu}$ from its spatial part $\chi_{k\ell}$ only.

In contrast to the fundamental response tensor, the conductivity tensor in Eq.~\eqref{eq_ohm} relates the spatial current to the observable electric field. Since a time-dependent electric field is in general also accompanied by a magnetic field, the effect of the latter is already contained in the microscopic conductivity tensor, which can hence be characterized as the total functional derivative (see \cite{ED1} for a discussion of this concept)
\begin{align}
& \sigma_{k\ell}(\vec x - \vec x'; t - t') = \frac{\de j_k(\vec x, t)}{\de E_\ell(\vec x', t')} \\[3pt]
& \equiv \frac{\delta j_k(\vec x, t)}{\delta E_\ell(\vec x', t')} + \int \! \de^3 \vec y \int \! \de s \, \frac{\delta j_k(\vec x, t)}{\delta B_j(\vec y, s)} \frac{\delta B_j(\vec y, s)}{\delta E_\ell(\vec x', t')}\,.
\end{align}
The thus defined conductivity tensor $\sigma_{k\ell}$ is related to the spatial part of the fundamental response tensor $\chi_{k\ell}$ by the standard relation (see e.g.~\cite{Giuliani, Nam})
\begin{equation}\label{eq_sigmachi}
\tsr\chi(\vec x - \vec x'; \omega) =\mathrm i \omega \h \tsr\sigma(\vec x - \vec x'; \omega) \,.
\end{equation}
Using this, we now find for the response of the spatial part of the current:
\begin{widetext}
\begin{align}
j_k(\vec x, \omega)&=\int \! \de^3 \vec x' \left\{ \chi_{k0}(\vec x - \vec x'; \omega) \, \frac{1}{c}  \, \varphi(\vec x', \omega)+\chi_{k\ell}(\vec x - \vec x'; \omega) A_\ell(\vec x', \omega) \right\}\label{eq_Ohm_derived4} \\[5pt]
&=\int \! \de^3 \vec x' \left\{ \left( \frac{1}{\mathrm i \omega} \frac{\partial}{\partial x'_\ell} \, \chi_{k\ell}(\vec x - \vec x'; \omega) \right) \varphi(\vec x', \omega)+\chi_{k\ell}(\vec x - \vec x'; \omega) \, A_\ell(\vec x', \omega) \right\} \\[5pt]
&=\frac{1}{\mathrm i \omega} \int\!\de^3 \vec x' \, \chi_{k\ell}(\vec x - \vec x'; \omega) \left\{ -\frac{\partial}{\partial x'_\ell} \, \varphi(\vec x', \omega) +\mathrm i \omega A_\ell(\vec x', \omega) \right\} \\[7pt]
&=\int \! \de^3 \vec x' \, \sigma_{k\ell}(\vec x - \vec x'; \omega) \h E_\ell(\vec x', \omega) \,, \label{eq_Ohm_derived}
\end{align}
\end{widetext}
which obviously coincides with the microscopic Ohm law. 
Besides proving Ohm's law from the more fundamental response law \eqref{eq_genResplaw}, the above calculation also provides a fully relativistic and gauge independent derivation of the universal relation \eqref{eq_sigmachi}. A similar calculation leads to the response law for the charge density
\begin{equation}
\rho(\vec x, \omega)=\frac{1}{\mathrm i \omega} \h \int \! \de^3 \vec x' \h \frac{\partial}{\partial x_k} \h \sigma_{k\ell}(\vec x - \vec x'; \omega) \h E_{\ell}(\vec x', \omega) \,,
\end{equation}
which is also consistent with the continuity equation for the induced four-current. We have thus shown that Ohm's law can be derived from a
covariant response theory. The seemingly paradoxical result that the induced current can be completely expressed in terms of the 
applied electric field (such that the magnetic field does not explicitly enter the description) 
stems from the fact that the conductivity tensor corresponds to a total functional derivative with respect to 
the external electric field \footnote{Strictly speaking, our derivation \eqref{eq_Ohm_derived4}--\eqref{eq_Ohm_derived}
will run into difficulties if there are static magnetic fields present. These would be proportional
to $\delta(\omega)$ in the frequency domain, and hence any formal manipulation with terms like $1/\omega$ would lead to singular results.
Often, however, the influence of static magnetic fields on the conduction current is simply neglected. If this approximation is untenable,
then the problem is not so much that Ohm's law looses its covariance. Instead, it looses its very validity giving way to a more general expansion
of the induced current in terms of the external fields. We refer the interested reader to the discussion in \cite[Sec.~6.7]{ED1}.}.

{\itshape Transformation law for the conductivity tensor.}---As we have derived Ohm's law from the fundamental, Lorentz-covariant response relation \eqref{eq_genResplaw}, it follows that Ohm's law holds in every inertial frame. Of course, it is understood that the conductivity tensor
itself obeys a transformation law, exactly as the fundamental response tensor \eqref{eq_fundconstr}.
For deriving this transformation law, we consider a general Lorentz transformation
$x'=\Lambda x$, where $\Lambda\in\rm{O}(1, 3)$, and assume that we are given the conductivity tensor $\sigma_{ij}(\vec k, \omega)$ in the unprimed coordinate system. 
The conductivity tensor in the primed coordinate system can then be derived in three steps as follows (see Fig.~\ref{fig_square}; a similar argument can already be found in Ref.~\cite{Melrose73}):
(i) By means of Eq.~\eqref{eq_sigmachi} one obtains the
spatial part $\chi_{ij}(\vec k, \omega)$ of the fundamental response tensor from the conductivity tensor, and by the constraints \eqref{eq_constr1}--\eqref{eq_constr3} one reconstructs from this the whole
fundamental response tensor $\chi\indices{^\mu_\nu}(\vec k, \omega) \equiv \chi(\vec k, \omega)$. (ii) The fundamental response tensor is a second-rank {\itshape Lorentz tensor}, hence it transforms according to
\begin{equation}\label{eq_trafo}
\chi'(\vec k',\omega') = \Lambda \h \chi(\vec k,\omega) \h \Lambda\inv \,, \qquad k' = \Lambda \h k \,.
\end{equation}
(iii) In the primed coordinate system, one invokes again Eq.~\eqref{eq_sigmachi}
to read out the conductivity tensor $\sigma'_{ij}(\vec k', \omega')$. The concatenation of these operations
then leads to a complicated (i.e.~non-tensorial) transformation law for the microscopic conductivity tensor under general Lorentz transformations.

Before deriving this transformation law explicitly, we note that 
under spatial rotations the conductivity tensor transforms according to
\begin{equation} \label{eq_tr_rot}
 \tsr\sigma{}'\big(\vec k', \omega\big) = \tsr R \, \tsr\sigma(\vec k, \omega) \h \tsr R{}^{-1}\,, \quad  \vec k' = \tsr R \h \vec k \,,
\end{equation}
where $\tsr R \in \rm{O}(3)$. For this reason, we call $\sigma_{k\ell}$ a second-rank {\itshape cartesian tensor}. On the other hand, every proper, ortho\-{}chronous Lorentz transformation can be factorized into a spatial rotation and a boost \cite{ScharfQED, Scheck}, while an arbitrary Lorentz transformation amounts to a proper, ortho\-{}chronous Lorentz transformation possibly combined with a time reversal and/or a parity transformation \cite{Kleinert, Streater}. Therefore, it suffices to study the transformation properties of the conductivity tensor under boosts, which have the general form
\begin{equation}
\Lambda(\vec v)= \left( \begin{array}{cc} \gamma &  -\gamma \vec v^{\rm T} \! / c \\[5pt] -\gamma \vec v / c & \tsr \Lambda \end{array} \right) \,. \smallskip
\end{equation}
Here, $\vec v$ is the velocity of the primed coordinate frame relative to the unprimed frame,
$\gamma = 1/\sqrt{1-|\vec v|^2/c^2}$, and
\begin{equation}
\tsr\Lambda = \tsr 1 +(\gamma-1)\h\frac{\vec v\vec v^{\rm T}}{|\vec v|^2} \,.
\end{equation}
In particular, the momenta and frequencies transform as
\begin{equation}
 \vec k' = \tsr \Lambda \vec k - \frac{\gamma \omega \vec v}{c^2} \,, \quad \ \omega' = \gamma(\omega - \vec v \! \cdot \! \vec k) \,. \label{eq_op}
\end{equation}
From Eq.~\eqref{eq_trafo} we obtain after some algebra the relation between the spatial components
\begin{equation}
\tsr\chi{}'(\vec k', \omega') = \tsr \Lambda \, \bigg( \tsr 1 - \frac{\vec v \vec k^{\rm T}}{\omega} \bigg) \, \tsr \chi(\vec k, \omega) \, \bigg( \tsr 1 - \frac{\vec k \vec v^{\rm T}}{\omega} \bigg) \, \tsr \Lambda \,.
\end{equation}
By applying Eq.~\eqref{eq_sigmachi} in both coordinate systems and taking into account the resulting factor $\omega / \omega'$ through Eq.~\eqref{eq_op},
we obtain the desired transformation law for the conductivity tensor:
\begin{widetext}
\begin{equation} \label{eq_reltrf}
\tsr\sigma{}'(\vec k', \omega') = \frac{1}{\gamma} \, \bigg( 1 - \frac{\vec v \! \cdot \! \vec k}{\omega} \bigg)^{\!\!-1} \,\h \tsr \Lambda \,\h \bigg( \tsr 1 - \frac{\vec v \vec k^{\rm T}}{\omega} \bigg) \,\h \tsr \sigma(\vec k, \omega) \,\h \bigg( \tsr 1 - \frac{\vec k \vec v^{\rm T}}{\omega} \bigg) \,\h \tsr \Lambda \,.
\end{equation}
In the following, we will also need the converse relation,
\begin{align}
\tsr\sigma{}(\vec k, \omega) & = \gamma \, \bigg( 1 - \frac{\vec v \! \cdot \! \vec k}{\omega} \bigg) \, \bigg( \tsr 1 - \frac{\vec v \vec k^{\rm T}}{\omega} \bigg)^{\!\!-1} \,\h \tsr \Lambda{}^{-1} \,\h \tsr \sigma{}'(\vec k', \omega') \,\h \tsr \Lambda{}^{-1} \,\h \bigg( \tsr 1 - \frac{\vec k \vec v^{\rm T}}{\omega} \bigg)^{\!\!-1} \\[8pt]
& = \gamma \, \bigg( \tsr 1 - \frac{\vec v \vec k^{\rm T}}{\omega} \bigg)^{\!\!-1} \,\h \tsr \Lambda{}^{-1} \,\h \tsr\sigma{}'(\vec k', \omega') \,\h \tsr \Lambda{}^{-1} \,\h \bigg( \bigg( 1 - \frac{\vec v \! \cdot \! \vec k}{\omega} \bigg) \h \tsr 1 + \frac{\vec k\vec v^{\rm T}}{\omega} \bigg) \,, \label{eq_reltrf_inv}
\end{align}
\end{widetext}
where in the last step we have used the identity
\begin{equation}
\bigg( \tsr 1 - \frac{\vec k \vec v^{\rm T}}{\omega} \bigg)^{\!\!-1} = \tsr 1 + \frac{\vec k \vec v^{\rm T}}{\omega - \vec v \! \cdot \! \vec k} \,.
\end{equation}
Note that Eq.~\eqref{eq_reltrf} represents a non-tensorial transformation law in contrast to the ordinary transformation law \eqref{eq_tr_rot} for rotations.

{\itshape Comparison with the standard textbook generalization of Ohm's law.}---Let us assume that the conductivity tensor is known in one particular inertial frame, which we identify with the primed coordinate system. Typically, this may be associated with the rest frame of the material probe (see, however, the discussion in the last section). Then Eq.~\eqref{eq_reltrf_inv} allows us to calculate the conductivity tensor in any inertial frame with constant velocity $-\vec v$ relative to the primed coordinate system. In particular, we can reexpress 
Ohm's law in the unprimed coordinate system,
$
 j_i(\vec k, \omega) = \sigma_{ij}(\vec k, \omega) E_j(\vec k, \omega),
$
in terms of the conductivity tensor of the primed coordinate system:
\begin{align}
 \vec j(\vec k, \omega) & = \gamma \, \bigg( \tsr 1 - \frac{\vec v \vec k^{\rm T}}{\omega} \bigg)^{\!\!-1} \,\h \tsr \Lambda{}^{-1} \,\h \tsr\sigma{}'(\vec k', \omega') \,\h \tsr \Lambda{}^{-1} \nonumber \\[3pt]
 & \quad \, \times \bigg( \bigg( 1 - \frac{\vec v \! \cdot \! \vec k}{\omega} \bigg) \h \tsr 1 + \frac{\vec k\vec v^{\rm T}}{\omega} \bigg) \, \vec E(\vec k, \omega) \,.
\end{align}
To simplify this expression, we use that
\begin{equation}
 \bigg( \tsr 1 - \frac{\vec v \vec k^{\rm T}}{\omega} \bigg) \, \vec j(\vec k, \omega) = \vec j(\vec k, \omega) - \vec v \rho(\vec k,\omega)
\end{equation}
by the continuity equation, and moreover,
\begin{align}
 & \left( 1 - \frac{\vec v \! \cdot \! \vec k}{\omega} \right) \vec E(\vec k, \omega) + \vec k \, \frac{\vec v \cdot \vec  E(\vec k, \omega)}{\omega} \nonumber \\[3pt]
 & = \vec E(\vec k, \omega) + \frac{\vec v \times \big( \vec k \times \vec E(\vec k, \omega) \big)}{\omega} \\[5pt]
 & = \vec E(\vec k, \omega) + \vec v \times \vec B(\vec k, \omega) \,,
\end{align}
where we have employed Faraday's law. Thus we obtain
\begin{align}
 \vec j(\vec k, \omega) - \vec v \rho(\vec k, \omega) & = \gamma \, \tsr \Lambda{}^{-1} \,\h \tsr\sigma{}'(\vec k', \omega') \,\h \tsr \Lambda{}^{-1}  \nonumber \\[5pt]
 & \quad \, \times \Big( \vec E(\vec k, \omega) + \vec v \times \vec B(\vec k, \omega) \Big) \,. \label{eq_relohm}
\end{align}
Now consider the special case where the conductivity of the primed coordinate system is scalar and constant, i.e.,
\begin{equation}
\sigma_{ij}'(\vec k', \omega') = \delta_{ij} \h \sigma' \,.
\end{equation}
Then Eq.~\eqref{eq_relohm} simplifies as
\begin{equation}
 \vec j - \vec v \rho = \gamma \h \sigma' \, \tsr \Lambda{}^{-2} \,
( \vec E + \vec v \times \vec B ) \,.
\end{equation}
Using that $(\gamma - 1)(\gamma + 1) = \gamma^2 |\vec v|^2 / c^2$, we find
\begin{equation}
\tsr \Lambda{}^{-2}
= \bigg( \tsr 1 + \frac{1 - \gamma}{\gamma} \, \frac{\vec v \vec v^{\rm T}}{|\vec v|^2}\bigg)^{\!\!2} \\[3pt]
= \tsr 1 - \frac{\vec v \vec v^{\rm T}}{c^2} \,,
\end{equation}
and consequently,
\begin{align}
 & \vec j - \rho \vec v = \gamma \h \sigma' \left( \vec E - \frac{\vec v}{c} \left( \frac{\vec v}{c} \cdot \vec E \right) + \vec v \times \vec B \right) \,. \label{eq_ohmapprox}
\end{align}
This formula is usually referred to as the relativistic generalization of Ohm's law in the textbook literature (see e.g.~\cite[Section~5.3]{RebhanRel}, \cite[Problem 11.16]{Jackson} and \cite[Problem 9-15]{Tsang}). We have shown that it is a special case of Eq.~\eqref{eq_relohm}, which in turn is a direct consequence of the relativistic transformation law for the conductivity tensor \eqref{eq_reltrf}.
Finally, we remark that in the limit where terms of order $|\vec v|^2/c^2$ are neglected, 
Eq.~\eqref{eq_ohmapprox} reduces to
\begin{equation} \label{st}
 \vec j - \vec v \rho =
 \sigma' ( \vec E + \vec v \times \vec B ) \,,
\end{equation}
which is also a standard textbook generalization of Ohm's law \cite{Tsang}. In fact, this last equation is obviously equivalent to $\vec j' = \sigma' \vec E'$, provided one uses the transformation laws $\vec j' = \vec j - \vec v \rho$ and $\vec E' = \vec E + \vec v \times \vec B$, respectively. Note, however, that these transformation laws for $\vec j$ and for $\vec E$ correspond to the electric limit and to the magnetic limit of the Maxwell equations, respectively \cite{Bellac}. Hence, these two transformation laws cannot be combined into a consistent non-relativistic limit of the Maxwell equations~\cite{Rousseaux}, and this shows that the generalized Ohm law is a non-trivial result which cannot be derived by elementary symmetry considerations.

{\itshape Paradoxes of the standard interpretation.}---Finally, in order to illustrate more clearly the advantages of our approach, let us describe some paradoxes of the standard interpretation of the relativistic Ohm law and their resolution by our approach. The standard interpretation (see e.g.~\cite{RebhanRel, Jackson, Tsang}) assumes that Ohm's law in the form \eqref{eq_naive_Ohm} holds in only one preferred inertial frame, while in all other inertial frames the generalized Ohm law \eqref{eq_ohmapprox} has to be employed. The latter equation contains in the form of the velocity parameter $\vec v$ a quantity which is neither associated with the material probe itself nor with the external perturbation in a given inertial frame. Therefore, 
given the electric and magnetic fields
in a certain inertial frame, one is not able to determine the induced current by means of Eq.~\eqref{eq_ohmapprox}. Instead, in the first place one would have to determine
the velocity $\vec v$ of the given inertial frame with respect to the preferred inertial frame where Ohm's law holds in the form \eqref{eq_naive_Ohm}.
This of course raises the question in which inertial frame Ohm's law in the form \eqref{eq_naive_Ohm} actually holds true. Here, it is tempting
to identify this inertial frame with the {\itshape rest frame of the center of mass} of the material probe. In actual fact, however, this answer is unsatisfactory as it leads to the following problems:

(i) First of all, it is not clear whose center of mass one should refer to. In the case of a solid crystal, for example, the center of mass will be determined almost exclusively by the positions of the nuclei, which have a much larger mass than the electrons. 
On the other hand, the conductivity of the sample will usually be determined by the electronic subsystem, or more precisely, by the conduction electrons. It is not clear, therefore, why the center of mass of the whole material probe should enter Ohm's law at all.
(ii) The center of mass of the electronic subsystem, on the other hand, will generally not be at rest. Instead, it will be accelerated by the external electric and magnetic fields. This would lead to additional complications in the applicability of Ohm's law in the form \eqref{eq_naive_Ohm}. 
(iii) Finally, on a microscopic level Ohm's law---like any other response law---relates the effect (i.e.~the induced current) at one space-time point $x$ to the perturbation (i.e.~the external electric field) at another space-time point $x'$ as described by Eq.~\eqref{eq_ohm}.
 The fundamental principle of {\itshape causality} now implies that the conductivity has to be retarded, i.e., the current at some space-time point~$x$ 
 can only be affected by the electric field at those space-time points $x'$ which lie in the backward light-cone of the point $x$. On the other hand, the center of mass is determined by integrating the mass density of the whole sample at a fixed time. If, for example, in the case of a relativistic plasma some far distant part of the sample started accelerating, then this would instantaneously affect also the center of mass of the whole sample. It is clear, therefore, that  the center of mass cannot enter any relativistic response law, because this would violate the principle of causality.

In our approach, these problems of the standard interpretation are naturally resolved, because Ohm's law in the form \eqref{eq_ohm}, if at all, 
actually holds in {\itshape any} intertial frame irrespectively of the center of mass. On the other hand, the conductivity tensor is not regarded as a mere number. Instead, its wavevector and frequency dependence as well as its relativistic transformation law are explicitly taken into account.

{\itshape Conclusion.}---Starting from the Lorentz-covariant microscopic response relation \eqref{eq_genResplaw},
we have established the validity of Ohm's law in every inertial frame. The reason for this unexpected covariance lies in the fact that the
conductivity tensor itself obeys a complicated (non-tensorial) transformation law, which we have derived explicitly in Eq.~\eqref{eq_reltrf}. 
Such a non-tensorial transformation law is of a particular interest, because it demonstrates that even an equation which does not involve a single Lorentz tensor can still be Lorentz covariant. Given the conductivity tensor in any particular intertial frame (such as the rest frame of the medium), our formula \eqref{eq_reltrf} can be used to compute the conductivity tensor in any other inertial frame. Thus, it provides a direct link to the experiment, where the relativistic current response is not only relevant for plasmas in large-scale astrophysics, but also for tabletop experiments with moving media in condensed matter and optical physics.
\begin{acknowledgments}
This research was supported by the Austrian Science Fund (FWF) within the SFB ViCoM, 
Grant No.~F41, and by the DFG Reseach Unit FOR 723. R.\,S.~thanks the Institute for Theoretical Physics at the TU Bergakademie Freiberg for its hospitality.
\end{acknowledgments}

\bibliography{/home/schober/Ronald/masterbib}

\begin{thebibliography}{62}%
\makeatletter
\providecommand \@ifxundefined [1]{%
 \@ifx{#1\undefined}
}%
\providecommand \@ifnum [1]{%
 \ifnum #1\expandafter \@firstoftwo
 \else \expandafter \@secondoftwo
 \fi
}%
\providecommand \@ifx [1]{%
 \ifx #1\expandafter \@firstoftwo
 \else \expandafter \@secondoftwo
 \fi
}%
\providecommand \natexlab [1]{#1}%
\providecommand \enquote  [1]{``#1''}%
\providecommand \bibnamefont  [1]{#1}%
\providecommand \bibfnamefont [1]{#1}%
\providecommand \citenamefont [1]{#1}%
\providecommand \href@noop [0]{\@secondoftwo}%
\providecommand \href [0]{\begingroup \@sanitize@url \@href}%
\providecommand \@href[1]{\@@startlink{#1}\@@href}%
\providecommand \@@href[1]{\endgroup#1\@@endlink}%
\providecommand \@sanitize@url [0]{\catcode `\\12\catcode `\$12\catcode
  `\&12\catcode `\#12\catcode `\^12\catcode `\_12\catcode `\%12\relax}%
\providecommand \@@startlink[1]{}%
\providecommand \@@endlink[0]{}%
\providecommand \url  [0]{\begingroup\@sanitize@url \@url }%
\providecommand \@url [1]{\endgroup\@href {#1}{\urlprefix }}%
\providecommand \urlprefix  [0]{URL }%
\providecommand \Eprint [0]{\href }%
\providecommand \doibase [0]{http://dx.doi.org/}%
\providecommand \selectlanguage [0]{\@gobble}%
\providecommand \bibinfo  [0]{\@secondoftwo}%
\providecommand \bibfield  [0]{\@secondoftwo}%
\providecommand \translation [1]{[#1]}%
\providecommand \BibitemOpen [0]{}%
\providecommand \bibitemStop [0]{}%
\providecommand \bibitemNoStop [0]{.\EOS\space}%
\providecommand \EOS [0]{\spacefactor3000\relax}%
\providecommand \BibitemShut  [1]{\csname bibitem#1\endcsname}%
\let\auto@bib@innerbib\@empty
\bibitem [{\citenamefont {Ohm}(1827)}]{Ohm1827}%
  \BibitemOpen
  \bibfield  {author} {\bibinfo {author} {\bibfnamefont {G.~S.}\ \bibnamefont
  {Ohm}},\ }\href@noop {} {\emph {\bibinfo {title} {{\itshape Die galvanische
  Kette, mathematisch bearbeitet}}}}\ (\bibinfo  {publisher} {T. H. Riemann},\
  \bibinfo {address} {Berlin},\ \bibinfo {year} {1827})\BibitemShut {NoStop}%
\bibitem [{\citenamefont {Morse}(1855)}]{Morse}%
  \BibitemOpen
  \bibfield  {author} {\bibinfo {author} {\bibfnamefont {S.~F.~B.}\
  \bibnamefont {Morse}},\ }in\ \href@noop {} {\emph {\bibinfo {booktitle}
  {{\itshape Shaffner's telegraph companion}}}},\ Vol.~\bibinfo {volume} {2},\
  \bibinfo {editor} {edited by\ \bibinfo {editor} {\bibfnamefont {T.~P.}\
  \bibnamefont {Shaffner}}}\ (\bibinfo  {publisher} {Pudney \& Russell},\
  \bibinfo {address} {New York},\ \bibinfo {year} {1855})\ pp.\ \bibinfo
  {pages} {6--96}\BibitemShut {NoStop}%
\bibitem [{\citenamefont {Butrica}(1990)}]{Butrica}%
  \BibitemOpen
  \bibfield  {author} {\bibinfo {author} {\bibfnamefont {A.~J.}\ \bibnamefont
  {Butrica}},\ }in\ \href@noop {} {\emph {\bibinfo {booktitle} {{\itshape
  Beyond history of science: essays in honor of Robert E. Schofield}}}},\
  \bibinfo {editor} {edited by\ \bibinfo {editor} {\bibfnamefont
  {E.}~\bibnamefont {Garber}}}\ (\bibinfo  {publisher} {Associated University
  Presses, Inc.},\ \bibinfo {address} {Cranbury, NJ},\ \bibinfo {year} {1990})\
  pp.\ \bibinfo {pages} {204--219}\BibitemShut {NoStop}%
\bibitem [{\citenamefont {Hodgkin}\ and\ \citenamefont
  {Huxley}(1952)}]{Hodgkin}%
  \BibitemOpen
  \bibfield  {author} {\bibinfo {author} {\bibfnamefont {A.~L.}\ \bibnamefont
  {Hodgkin}}\ and\ \bibinfo {author} {\bibfnamefont {A.~F.}\ \bibnamefont
  {Huxley}},\ }\href@noop {} {\bibfield  {journal} {\bibinfo  {journal} {J.
  Physiol.}\ }\textbf {\bibinfo {volume} {{\bfseries 117}}},\ \bibinfo {pages}
  {500} (\bibinfo {year} {1952})}\BibitemShut {NoStop}%
\bibitem [{\citenamefont {Lytton}\ and\ \citenamefont {Kerr}(2013)}]{Lytton}%
  \BibitemOpen
  \bibfield  {author} {\bibinfo {author} {\bibfnamefont {W.~W.}\ \bibnamefont
  {Lytton}}\ and\ \bibinfo {author} {\bibfnamefont {C.~C.}\ \bibnamefont
  {Kerr}},\ }in\ \href@noop {} {\emph {\bibinfo {booktitle} {{\itshape
  Neuroscience in the 21st century: from basic to clinical}}}},\ \bibinfo
  {editor} {edited by\ \bibinfo {editor} {\bibfnamefont {D.~W.}\ \bibnamefont
  {Pfaff}}}\ (\bibinfo  {publisher} {Springer Science+Business Media, LLC},\
  \bibinfo {address} {New York},\ \bibinfo {year} {2013})\ pp.\ \bibinfo
  {pages} {2275--2299}\BibitemShut {NoStop}%
\bibitem [{\citenamefont {Parikh}\ and\ \citenamefont
  {Wilczek}(1998)}]{Parikh}%
  \BibitemOpen
  \bibfield  {author} {\bibinfo {author} {\bibfnamefont {M.~K.}\ \bibnamefont
  {Parikh}}\ and\ \bibinfo {author} {\bibfnamefont {F.}~\bibnamefont
  {Wilczek}},\ }\href@noop {} {\bibfield  {journal} {\bibinfo  {journal} {Phys.
  Rev. D}\ }\textbf {\bibinfo {volume} {{\bfseries 58}}},\ \bibinfo {pages}
  {064011} (\bibinfo {year} {1998})}\BibitemShut {NoStop}%
\bibitem [{\citenamefont {Thorne}\ \emph {et~al.}(1986)\citenamefont {Thorne},
  \citenamefont {Price},\ and\ \citenamefont {{MacDonald}}}]{Thorne}%
  \BibitemOpen
  \bibinfo {editor} {\bibfnamefont {K.~S.}\ \bibnamefont {Thorne}}, \bibinfo
  {editor} {\bibfnamefont {R.~H.}\ \bibnamefont {Price}}, \ and\ \bibinfo
  {editor} {\bibfnamefont {D.~A.}\ \bibnamefont {{MacDonald}}},\ eds.,\
  \href@noop {} {\emph {\bibinfo {title} {{\itshape Black holes: the membrane
  paradigm}}}}\ (\bibinfo  {publisher} {Yale University Press},\ \bibinfo
  {year} {1986})\BibitemShut {NoStop}%
\bibitem [{\citenamefont {Hartnoll}\ and\ \citenamefont
  {Herzog}(2007)}]{String}%
  \BibitemOpen
  \bibfield  {author} {\bibinfo {author} {\bibfnamefont {S.~A.}\ \bibnamefont
  {Hartnoll}}\ and\ \bibinfo {author} {\bibfnamefont {C.~P.}\ \bibnamefont
  {Herzog}},\ }\href@noop {} {\bibfield  {journal} {\bibinfo  {journal} {Phys.
  Rev. D}\ }\textbf {\bibinfo {volume} {{\bfseries 76}}},\ \bibinfo {pages}
  {106012} (\bibinfo {year} {2007})}\BibitemShut {NoStop}%
\bibitem [{\citenamefont {Weber}\ \emph {et~al.}(2012)\citenamefont {Weber},
  \citenamefont {Mahapatra}, \citenamefont {Ryu}, \citenamefont {Lee},
  \citenamefont {Fuhrer}, \citenamefont {Reusch}, \citenamefont {Thompson},
  \citenamefont {Lee}, \citenamefont {Klimeck}, \citenamefont {Hollenberg},\
  and\ \citenamefont {Simmons}}]{Weber}%
  \BibitemOpen
  \bibfield  {author} {\bibinfo {author} {\bibfnamefont {B.}~\bibnamefont
  {Weber}}, \bibinfo {author} {\bibfnamefont {S.}~\bibnamefont {Mahapatra}},
  \bibinfo {author} {\bibfnamefont {H.}~\bibnamefont {Ryu}}, \bibinfo {author}
  {\bibfnamefont {S.}~\bibnamefont {Lee}}, \bibinfo {author} {\bibfnamefont
  {A.}~\bibnamefont {Fuhrer}}, \bibinfo {author} {\bibfnamefont {T.~C.~G.}\
  \bibnamefont {Reusch}}, \bibinfo {author} {\bibfnamefont {D.~L.}\
  \bibnamefont {Thompson}}, \bibinfo {author} {\bibfnamefont {W.~C.~T.}\
  \bibnamefont {Lee}}, \bibinfo {author} {\bibfnamefont {G.}~\bibnamefont
  {Klimeck}}, \bibinfo {author} {\bibfnamefont {L.~C.~L.}\ \bibnamefont
  {Hollenberg}}, \ and\ \bibinfo {author} {\bibfnamefont {M.~Y.}\ \bibnamefont
  {Simmons}},\ }\href@noop {} {\bibfield  {journal} {\bibinfo  {journal}
  {Science}\ }\textbf {\bibinfo {volume} {{\bfseries 335}}},\ \bibinfo {pages}
  {64} (\bibinfo {year} {2012})}\BibitemShut {NoStop}%
\bibitem [{\citenamefont {Ryu}\ \emph {et~al.}(2013)\citenamefont {Ryu},
  \citenamefont {Lee}, \citenamefont {Weber}, \citenamefont {Mahapatra},
  \citenamefont {Hollenberg}, \citenamefont {Simmons},\ and\ \citenamefont
  {Klimeck}}]{Ryu}%
  \BibitemOpen
  \bibfield  {author} {\bibinfo {author} {\bibfnamefont {H.}~\bibnamefont
  {Ryu}}, \bibinfo {author} {\bibfnamefont {S.}~\bibnamefont {Lee}}, \bibinfo
  {author} {\bibfnamefont {B.}~\bibnamefont {Weber}}, \bibinfo {author}
  {\bibfnamefont {S.}~\bibnamefont {Mahapatra}}, \bibinfo {author}
  {\bibfnamefont {L.~C.~L.}\ \bibnamefont {Hollenberg}}, \bibinfo {author}
  {\bibfnamefont {M.~Y.}\ \bibnamefont {Simmons}}, \ and\ \bibinfo {author}
  {\bibfnamefont {G.}~\bibnamefont {Klimeck}},\ }\href@noop {} {\bibfield
  {journal} {\bibinfo  {journal} {Nanoscale}\ }\textbf {\bibinfo {volume}
  {{\bfseries 5}}},\ \bibinfo {pages} {8666} (\bibinfo {year}
  {2013})}\BibitemShut {NoStop}%
\bibitem [{\citenamefont {Ferry}(2008)}]{Ferry}%
  \BibitemOpen
  \bibfield  {author} {\bibinfo {author} {\bibfnamefont {D.~K.}\ \bibnamefont
  {Ferry}},\ }\href@noop {} {\bibfield  {journal} {\bibinfo  {journal}
  {Science}\ }\textbf {\bibinfo {volume} {{\bfseries 319}}},\ \bibinfo {pages}
  {579} (\bibinfo {year} {2008})}\BibitemShut {NoStop}%
\bibitem [{\citenamefont {Fuechsle}\ \emph {et~al.}(2012)\citenamefont
  {Fuechsle}, \citenamefont {Miwa}, \citenamefont {Mahapatra}, \citenamefont
  {Ryu}, \citenamefont {Lee}, \citenamefont {Warschkow}, \citenamefont
  {Hollenberg}, \citenamefont {Klimeck},\ and\ \citenamefont
  {Simmons}}]{Fuechsle}%
  \BibitemOpen
  \bibfield  {author} {\bibinfo {author} {\bibfnamefont {M.}~\bibnamefont
  {Fuechsle}}, \bibinfo {author} {\bibfnamefont {J.~A.}\ \bibnamefont {Miwa}},
  \bibinfo {author} {\bibfnamefont {S.}~\bibnamefont {Mahapatra}}, \bibinfo
  {author} {\bibfnamefont {H.}~\bibnamefont {Ryu}}, \bibinfo {author}
  {\bibfnamefont {S.}~\bibnamefont {Lee}}, \bibinfo {author} {\bibfnamefont
  {O.}~\bibnamefont {Warschkow}}, \bibinfo {author} {\bibfnamefont {L.~C.~L.}\
  \bibnamefont {Hollenberg}}, \bibinfo {author} {\bibfnamefont
  {G.}~\bibnamefont {Klimeck}}, \ and\ \bibinfo {author} {\bibfnamefont
  {M.~Y.}\ \bibnamefont {Simmons}},\ }\href@noop {} {\bibfield  {journal}
  {\bibinfo  {journal} {Nat. Nanotechnol.}\ }\textbf {\bibinfo {volume}
  {{\bfseries 7}}},\ \bibinfo {pages} {242} (\bibinfo {year}
  {2012})}\BibitemShut {NoStop}%
\bibitem [{\citenamefont {Peierls}(1960)}]{Peierls}%
  \BibitemOpen
  \bibfield  {author} {\bibinfo {author} {\bibfnamefont {R.~E.}\ \bibnamefont
  {Peierls}},\ }in\ \href@noop {} {\emph {\bibinfo {booktitle} {{\itshape
  Theoretical physics in the twentieth century: a memorial volume to Wolfgang
  Pauli}}}},\ \bibinfo {editor} {edited by\ \bibinfo {editor} {\bibfnamefont
  {M.}~\bibnamefont {Fierz}}\ and\ \bibinfo {editor} {\bibfnamefont {V.~F.}\
  \bibnamefont {Weisskopf}}}\ (\bibinfo  {publisher} {Interscience
  Publishers},\ \bibinfo {address} {New York},\ \bibinfo {year} {1960})\ pp.\
  \bibinfo {pages} {140--160}\BibitemShut {NoStop}%
\bibitem [{\citenamefont {Chernov}\ \emph {et~al.}(1993)\citenamefont
  {Chernov}, \citenamefont {Eyink}, \citenamefont {Lebowitz},\ and\
  \citenamefont {Sinai}}]{Chernov}%
  \BibitemOpen
  \bibfield  {author} {\bibinfo {author} {\bibfnamefont {N.~I.}\ \bibnamefont
  {Chernov}}, \bibinfo {author} {\bibfnamefont {G.~L.}\ \bibnamefont {Eyink}},
  \bibinfo {author} {\bibfnamefont {J.~L.}\ \bibnamefont {Lebowitz}}, \ and\
  \bibinfo {author} {\bibfnamefont {Y.~G.}\ \bibnamefont {Sinai}},\ }\href@noop
  {} {\bibfield  {journal} {\bibinfo  {journal} {Phys. Rev. Lett.}\ }\textbf
  {\bibinfo {volume} {{\bfseries 70}}},\ \bibinfo {pages} {2209} (\bibinfo
  {year} {1993})}\BibitemShut {NoStop}%
\bibitem [{\citenamefont {Krall}\ and\ \citenamefont
  {Trivelpiece}(1973)}]{Krall}%
  \BibitemOpen
  \bibfield  {author} {\bibinfo {author} {\bibfnamefont {N.~A.}\ \bibnamefont
  {Krall}}\ and\ \bibinfo {author} {\bibfnamefont {A.~W.}\ \bibnamefont
  {Trivelpiece}},\ }\href@noop {} {\emph {\bibinfo {title} {{\itshape
  Principles of plasma physics}}}},\ \textnormal{International Series in Pure
  and Applied Physics}\ (\bibinfo  {publisher} {McGraw-Hill, Inc.},\ \bibinfo
  {address} {New York},\ \bibinfo {year} {1973})\BibitemShut {NoStop}%
\bibitem [{\citenamefont {Lorrain}\ \emph {et~al.}(2006)\citenamefont
  {Lorrain}, \citenamefont {Lorrain},\ and\ \citenamefont {Houle}}]{Lorrain}%
  \BibitemOpen
  \bibfield  {author} {\bibinfo {author} {\bibfnamefont {P.}~\bibnamefont
  {Lorrain}}, \bibinfo {author} {\bibfnamefont {F.}~\bibnamefont {Lorrain}}, \
  and\ \bibinfo {author} {\bibfnamefont {S.}~\bibnamefont {Houle}},\
  }\href@noop {} {\emph {\bibinfo {title} {{\itshape Magneto-fluid dynamics:
  fundamentals and case studies of natural phenomena}}}}\ (\bibinfo
  {publisher} {Springer Science+Business Media, LLC},\ \bibinfo {address} {New
  York},\ \bibinfo {year} {2006})\BibitemShut {NoStop}%
\bibitem [{\citenamefont {Lichnerowicz}(1967)}]{Lichnerowicz}%
  \BibitemOpen
  \bibfield  {author} {\bibinfo {author} {\bibfnamefont {A.}~\bibnamefont
  {Lichnerowicz}},\ }\href@noop {} {\emph {\bibinfo {title} {{\itshape
  Relativistic hydrodynamics and magnetohydrodynamics: lectures on the
  existence of solutions}}}},\ \textnormal{The Mathematical Physics Monograph
  Series}\ (\bibinfo  {publisher} {W. A. Benjamin, Inc.},\ \bibinfo {address}
  {New York},\ \bibinfo {year} {1967})\BibitemShut {NoStop}%
\bibitem [{\citenamefont {Palenzuela}\ \emph {et~al.}(2009)\citenamefont
  {Palenzuela}, \citenamefont {Lehner}, \citenamefont {Reula},\ and\
  \citenamefont {Rezzolla}}]{Palenzuela}%
  \BibitemOpen
  \bibfield  {author} {\bibinfo {author} {\bibfnamefont {C.}~\bibnamefont
  {Palenzuela}}, \bibinfo {author} {\bibfnamefont {L.}~\bibnamefont {Lehner}},
  \bibinfo {author} {\bibfnamefont {O.}~\bibnamefont {Reula}}, \ and\ \bibinfo
  {author} {\bibfnamefont {L.}~\bibnamefont {Rezzolla}},\ }\href@noop {}
  {\bibfield  {journal} {\bibinfo  {journal} {MNRAS}\ }\textbf {\bibinfo
  {volume} {{\bfseries 394}}},\ \bibinfo {pages} {1727} (\bibinfo {year}
  {2009})}\BibitemShut {NoStop}%
\bibitem [{\citenamefont {Meier}(2004)}]{Meier}%
  \BibitemOpen
  \bibfield  {author} {\bibinfo {author} {\bibfnamefont {D.~L.}\ \bibnamefont
  {Meier}},\ }\href@noop {} {\bibfield  {journal} {\bibinfo  {journal} {ApJ}\
  }\textbf {\bibinfo {volume} {{\bfseries 605}}},\ \bibinfo {pages} {340}
  (\bibinfo {year} {2004})}\BibitemShut {NoStop}%
\bibitem [{\citenamefont {Holcomb}\ and\ \citenamefont
  {Tajima}(1989)}]{Holcomb}%
  \BibitemOpen
  \bibfield  {author} {\bibinfo {author} {\bibfnamefont {K.~A.}\ \bibnamefont
  {Holcomb}}\ and\ \bibinfo {author} {\bibfnamefont {T.}~\bibnamefont
  {Tajima}},\ }\href@noop {} {\bibfield  {journal} {\bibinfo  {journal} {Phys.
  Rev. D}\ }\textbf {\bibinfo {volume} {{\bfseries 40}}},\ \bibinfo {pages}
  {3809} (\bibinfo {year} {1989})}\BibitemShut {NoStop}%
\bibitem [{\citenamefont {Ardavan}(1976)}]{Ardavan}%
  \BibitemOpen
  \bibfield  {author} {\bibinfo {author} {\bibfnamefont {H.}~\bibnamefont
  {Ardavan}},\ }\href@noop {} {\bibfield  {journal} {\bibinfo  {journal} {ApJ}\
  }\textbf {\bibinfo {volume} {{\bfseries 203}}},\ \bibinfo {pages} {226}
  (\bibinfo {year} {1976})}\BibitemShut {NoStop}%
\bibitem [{\citenamefont {Bekenstein}\ and\ \citenamefont
  {Oron}(1978)}]{Bekenstein}%
  \BibitemOpen
  \bibfield  {author} {\bibinfo {author} {\bibfnamefont {J.~D.}\ \bibnamefont
  {Bekenstein}}\ and\ \bibinfo {author} {\bibfnamefont {E.}~\bibnamefont
  {Oron}},\ }\href@noop {} {\bibfield  {journal} {\bibinfo  {journal} {Phys.
  Rev. D}\ }\textbf {\bibinfo {volume} {{\bfseries 18}}},\ \bibinfo {pages}
  {1809} (\bibinfo {year} {1978})}\BibitemShut {NoStop}%
\bibitem [{\citenamefont {Blackman}\ and\ \citenamefont
  {Field}(1993)}]{Blackman}%
  \BibitemOpen
  \bibfield  {author} {\bibinfo {author} {\bibfnamefont {E.~G.}\ \bibnamefont
  {Blackman}}\ and\ \bibinfo {author} {\bibfnamefont {G.~B.}\ \bibnamefont
  {Field}},\ }\href@noop {} {\bibfield  {journal} {\bibinfo  {journal} {Phys.
  Rev. Lett.}\ }\textbf {\bibinfo {volume} {{\bfseries 71}}},\ \bibinfo {pages}
  {3481} (\bibinfo {year} {1993})}\BibitemShut {NoStop}%
\bibitem [{\citenamefont {Gedalin}(1996)}]{Gedalin}%
  \BibitemOpen
  \bibfield  {author} {\bibinfo {author} {\bibfnamefont {M.}~\bibnamefont
  {Gedalin}},\ }\href@noop {} {\bibfield  {journal} {\bibinfo  {journal} {Phys.
  Rev. Lett.}\ }\textbf {\bibinfo {volume} {{\bfseries 76}}},\ \bibinfo {pages}
  {3340} (\bibinfo {year} {1996})}\BibitemShut {NoStop}%
\bibitem [{\citenamefont {Khanna}(1998)}]{Khanna}%
  \BibitemOpen
  \bibfield  {author} {\bibinfo {author} {\bibfnamefont {R.}~\bibnamefont
  {Khanna}},\ }\href@noop {} {\bibfield  {journal} {\bibinfo  {journal}
  {MNRAS}\ }\textbf {\bibinfo {volume} {{\bfseries 294}}},\ \bibinfo {pages}
  {673} (\bibinfo {year} {1998})}\BibitemShut {NoStop}%
\bibitem [{\citenamefont {Kremer}\ and\ \citenamefont {Patsko}(2003)}]{Kremer}%
  \BibitemOpen
  \bibfield  {author} {\bibinfo {author} {\bibfnamefont {G.~M.}\ \bibnamefont
  {Kremer}}\ and\ \bibinfo {author} {\bibfnamefont {C.~H.}\ \bibnamefont
  {Patsko}},\ }\href@noop {} {\bibfield  {journal} {\bibinfo  {journal}
  {{Physica A}}\ }\textbf {\bibinfo {volume} {{\bfseries 322}}},\ \bibinfo
  {pages} {329} (\bibinfo {year} {2003})}\BibitemShut {NoStop}%
\bibitem [{\citenamefont {Kandus}\ and\ \citenamefont {Tsagas}(2008)}]{Kandus}%
  \BibitemOpen
  \bibfield  {author} {\bibinfo {author} {\bibfnamefont {A.}~\bibnamefont
  {Kandus}}\ and\ \bibinfo {author} {\bibfnamefont {C.~G.}\ \bibnamefont
  {Tsagas}},\ }\href@noop {} {\bibfield  {journal} {\bibinfo  {journal}
  {MNRAS}\ }\textbf {\bibinfo {volume} {{\bfseries 385}}},\ \bibinfo {pages}
  {883} (\bibinfo {year} {2008})}\BibitemShut {NoStop}%
\bibitem [{\citenamefont {Sherlock}(2010)}]{Sherlock}%
  \BibitemOpen
  \bibfield  {author} {\bibinfo {author} {\bibfnamefont {M.}~\bibnamefont
  {Sherlock}},\ }\href@noop {} {\bibfield  {journal} {\bibinfo  {journal}
  {Phys. Rev. Lett.}\ }\textbf {\bibinfo {volume} {{\bfseries 104}}},\ \bibinfo
  {pages} {205004} (\bibinfo {year} {2010})}\BibitemShut {NoStop}%
\bibitem [{\citenamefont {Ahmedov}(2011)}]{Ahmedov2011}%
  \BibitemOpen
  \bibfield  {author} {\bibinfo {author} {\bibfnamefont {B.~B.}\ \bibnamefont
  {Ahmedov}},\ }\href@noop {} {\bibfield  {journal} {\bibinfo  {journal}
  {Astrophys. Space Sci.}\ }\textbf {\bibinfo {volume} {{\bfseries 331}}},\
  \bibinfo {pages} {565} (\bibinfo {year} {2011})}\BibitemShut {NoStop}%
\bibitem [{\citenamefont {Bucciantini}\ and\ \citenamefont
  {Del~Zanna}(2013)}]{Bucciantini}%
  \BibitemOpen
  \bibfield  {author} {\bibinfo {author} {\bibfnamefont {N.}~\bibnamefont
  {Bucciantini}}\ and\ \bibinfo {author} {\bibfnamefont {L.}~\bibnamefont
  {Del~Zanna}},\ }\href@noop {} {\bibfield  {journal} {\bibinfo  {journal}
  {MNRAS}\ }\textbf {\bibinfo {volume} {{\bfseries 428}}},\ \bibinfo {pages}
  {71} (\bibinfo {year} {2013})}\BibitemShut {NoStop}%
\bibitem [{\citenamefont {Tsamparlis}(2010)}]{Tsamparlis}%
  \BibitemOpen
  \bibfield  {author} {\bibinfo {author} {\bibfnamefont {M.}~\bibnamefont
  {Tsamparlis}},\ }\href@noop {} {\emph {\bibinfo {title} {{\itshape Special
  relativity: an introduction with 200 problems and solutions}}}}\ (\bibinfo
  {publisher} {Springer-Verlag},\ \bibinfo {address} {Berlin/Heidelberg},\
  \bibinfo {year} {2010})\BibitemShut {NoStop}%
\bibitem [{\citenamefont {Leonhardt}\ and\ \citenamefont
  {Philbin}(2006)}]{Leonhardt}%
  \BibitemOpen
  \bibfield  {author} {\bibinfo {author} {\bibfnamefont {U.}~\bibnamefont
  {Leonhardt}}\ and\ \bibinfo {author} {\bibfnamefont {T.~G.}\ \bibnamefont
  {Philbin}},\ }\href@noop {} {\bibfield  {journal} {\bibinfo  {journal} {New
  J. Phys.}\ }\textbf {\bibinfo {volume} {{\bfseries 8}}},\ \bibinfo {pages}
  {247} (\bibinfo {year} {2006})}\BibitemShut {NoStop}%
\bibitem [{\citenamefont {Mackay}\ and\ \citenamefont
  {Lakhtakia}(2007)}]{Mackay07}%
  \BibitemOpen
  \bibfield  {author} {\bibinfo {author} {\bibfnamefont {T.~G.}\ \bibnamefont
  {Mackay}}\ and\ \bibinfo {author} {\bibfnamefont {A.}~\bibnamefont
  {Lakhtakia}},\ }\href@noop {} {\bibfield  {journal} {\bibinfo  {journal}
  {Proc. R. Soc. A}\ }\textbf {\bibinfo {volume} {{\bfseries 463}}},\ \bibinfo
  {pages} {397} (\bibinfo {year} {2007})}\BibitemShut {NoStop}%
\bibitem [{\citenamefont {Joos}(1959)}]{Joos}%
  \BibitemOpen
  \bibfield  {author} {\bibinfo {author} {\bibfnamefont {G.}~\bibnamefont
  {Joos}},\ }\href@noop {} {\emph {\bibinfo {title} {{\itshape Lehrbuch der
  Theoretischen Physik}}}}\ (\bibinfo  {publisher} {Akademische
  Verlagsgesellschaft Gees \& Portig K.-G.},\ \bibinfo {address} {Leipzig},\
  \bibinfo {year} {1959})\BibitemShut {NoStop}%
\bibitem [{\citenamefont {Bergmann}\ and\ \citenamefont
  {Schaefer}(1999)}]{Bergmann}%
  \BibitemOpen
  \bibfield  {author} {\bibinfo {author} {\bibfnamefont {L.}~\bibnamefont
  {Bergmann}}\ and\ \bibinfo {author} {\bibfnamefont {C.}~\bibnamefont
  {Schaefer}},\ }\href@noop {} {\emph {\bibinfo {title} {{\itshape Optics of
  waves and particles}}}}\ (\bibinfo  {publisher} {Walter de Gruyter},\
  \bibinfo {address} {Berlin},\ \bibinfo {year} {1999})\BibitemShut {NoStop}%
\bibitem [{\citenamefont {Laue}(1907)}]{Laue}%
  \BibitemOpen
  \bibfield  {author} {\bibinfo {author} {\bibfnamefont {M.}~\bibnamefont
  {Laue}},\ }\href@noop {} {\bibfield  {journal} {\bibinfo  {journal} {Ann.
  Phys. (Berlin)}\ }\textbf {\bibinfo {volume} {{\bfseries 328}}},\ \bibinfo
  {pages} {989} (\bibinfo {year} {1907})}\BibitemShut {NoStop}%
\bibitem [{\citenamefont {Cho}(2010)}]{Cho10}%
  \BibitemOpen
  \bibfield  {author} {\bibinfo {author} {\bibfnamefont {K.}~\bibnamefont
  {Cho}},\ }\href@noop {} {\emph {\bibinfo {title} {{\itshape Reconstruction of
  macroscopic Maxwell equations: a single susceptibility theory}}}},\ \bibinfo
  {series} {\textnormal{Springer Tracts in Modern Physics}}, Vol.\ \bibinfo
  {volume} {237}\ (\bibinfo  {publisher} {Springer-Verlag},\ \bibinfo {address}
  {Berlin/Heidelberg},\ \bibinfo {year} {2010})\BibitemShut {NoStop}%
\bibitem [{\citenamefont {Melrose}\ and\ \citenamefont
  {McPhedran}(1991)}]{Melrose}%
  \BibitemOpen
  \bibfield  {author} {\bibinfo {author} {\bibfnamefont {D.~B.}\ \bibnamefont
  {Melrose}}\ and\ \bibinfo {author} {\bibfnamefont {R.~C.}\ \bibnamefont
  {McPhedran}},\ }\href@noop {} {\emph {\bibinfo {title} {{\itshape
  Electromagnetic processes in dispersive media: a treatment based on the
  dielectric tensor}}}}\ (\bibinfo  {publisher} {Cambridge University Press},\
  \bibinfo {address} {Cambridge},\ \bibinfo {year} {1991})\BibitemShut
  {NoStop}%
\bibitem [{\citenamefont {Starke}\ and\ \citenamefont {Schober}(2015)}]{ED1}%
  \BibitemOpen
  \bibfield  {author} {\bibinfo {author} {\bibfnamefont {R.}~\bibnamefont
  {Starke}}\ and\ \bibinfo {author} {\bibfnamefont {G.~A.~H.}\ \bibnamefont
  {Schober}},\ }\href@noop {} {\bibfield  {journal} {\bibinfo  {journal} {Phot.
  Nano. Fund. Appl.}\ }\textbf {\bibinfo {volume} {{\bfseries 14}}},\ \bibinfo
  {pages} {1} (\bibinfo {year} {2015})},\ \bibinfo {note} {{See also
  arXiv:1401.6800 [cond-mat.mtrl-sci]}}\BibitemShut {NoStop}%
\bibitem [{\citenamefont {Liberman}\ and\ \citenamefont
  {Zangwill}(1984)}]{Liberman}%
  \BibitemOpen
  \bibfield  {author} {\bibinfo {author} {\bibfnamefont {D.~A.}\ \bibnamefont
  {Liberman}}\ and\ \bibinfo {author} {\bibfnamefont {A.}~\bibnamefont
  {Zangwill}},\ }\href@noop {} {\bibfield  {journal} {\bibinfo  {journal}
  {Comput. Phys. Commun.}\ }\textbf {\bibinfo {volume} {{\bfseries 32}}},\
  \bibinfo {pages} {75} (\bibinfo {year} {1984})}\BibitemShut {NoStop}%
\bibitem [{\citenamefont {Neutze}\ and\ \citenamefont
  {Stedman}(1998)}]{Neutze}%
  \BibitemOpen
  \bibfield  {author} {\bibinfo {author} {\bibfnamefont {R.}~\bibnamefont
  {Neutze}}\ and\ \bibinfo {author} {\bibfnamefont {G.~E.}\ \bibnamefont
  {Stedman}},\ }\href@noop {} {\bibfield  {journal} {\bibinfo  {journal} {Phys.
  Rev. A}\ }\textbf {\bibinfo {volume} {{\bfseries 58}}},\ \bibinfo {pages}
  {82} (\bibinfo {year} {1998})}\BibitemShut {NoStop}%
\bibitem [{\citenamefont {de~Boeij}\ \emph {et~al.}(2001)\citenamefont
  {de~Boeij}, \citenamefont {Kootstra},\ and\ \citenamefont
  {Snijders}}]{Boeij}%
  \BibitemOpen
  \bibfield  {author} {\bibinfo {author} {\bibfnamefont {P.~L.}\ \bibnamefont
  {de~Boeij}}, \bibinfo {author} {\bibfnamefont {F.}~\bibnamefont {Kootstra}},
  \ and\ \bibinfo {author} {\bibfnamefont {J.~G.}\ \bibnamefont {Snijders}},\
  }\href@noop {} {\bibfield  {journal} {\bibinfo  {journal} {Int. J. Quantum
  Chem.}\ }\textbf {\bibinfo {volume} {{\bfseries 85}}},\ \bibinfo {pages}
  {449} (\bibinfo {year} {2001})}\BibitemShut {NoStop}%
\bibitem [{\citenamefont {Romaniello}\ and\ \citenamefont
  {Boeij}(2005)}]{Romaniello}%
  \BibitemOpen
  \bibfield  {author} {\bibinfo {author} {\bibfnamefont {P.}~\bibnamefont
  {Romaniello}}\ and\ \bibinfo {author} {\bibfnamefont {P.~L.}\ \bibnamefont
  {Boeij}},\ }\href@noop {} {\bibfield  {journal} {\bibinfo  {journal} {J.
  Phys. Chem.}\ }\textbf {\bibinfo {volume} {{\bfseries 122}}},\ \bibinfo
  {pages} {164303} (\bibinfo {year} {2005})}\BibitemShut {NoStop}%
\bibitem [{\citenamefont {Lee}\ \emph {et~al.}(2011)\citenamefont {Lee},
  \citenamefont {Schober}, \citenamefont {Bahramy}, \citenamefont {Murakawa},
  \citenamefont {Onose}, \citenamefont {Arita}, \citenamefont {Nagaosa},\ and\
  \citenamefont {Tokura}}]{Lee}%
  \BibitemOpen
  \bibfield  {author} {\bibinfo {author} {\bibfnamefont {J.~S.}\ \bibnamefont
  {Lee}}, \bibinfo {author} {\bibfnamefont {G.~A.~H.}\ \bibnamefont {Schober}},
  \bibinfo {author} {\bibfnamefont {M.~S.}\ \bibnamefont {Bahramy}}, \bibinfo
  {author} {\bibfnamefont {H.}~\bibnamefont {Murakawa}}, \bibinfo {author}
  {\bibfnamefont {Y.}~\bibnamefont {Onose}}, \bibinfo {author} {\bibfnamefont
  {R.}~\bibnamefont {Arita}}, \bibinfo {author} {\bibfnamefont
  {N.}~\bibnamefont {Nagaosa}}, \ and\ \bibinfo {author} {\bibfnamefont
  {Y.}~\bibnamefont {Tokura}},\ }\href {\doibase
  10.1103/PhysRevLett.107.117401} {\bibfield  {journal} {\bibinfo  {journal}
  {Phys. Rev. Lett.}\ }\textbf {\bibinfo {volume} {{\bfseries 107}}},\ \bibinfo
  {pages} {117401} (\bibinfo {year} {2011})}\BibitemShut {NoStop}%
\bibitem [{\citenamefont {Keshav}\ \emph {et~al.}(2011)\citenamefont {Keshav},
  \citenamefont {Lee}, \citenamefont {Rishi}, \citenamefont {Mohite},
  \citenamefont {Galande}, \citenamefont {Ajayan}, \citenamefont {Dattelbaum},
  \citenamefont {Htoon}, \citenamefont {Taylor},\ and\ \citenamefont
  {Prasankumar}}]{Dani}%
  \BibitemOpen
  \bibfield  {author} {\bibinfo {author} {\bibfnamefont {D.~M.}\ \bibnamefont
  {Keshav}}, \bibinfo {author} {\bibfnamefont {J.}~\bibnamefont {Lee}},
  \bibinfo {author} {\bibfnamefont {S.}~\bibnamefont {Rishi}}, \bibinfo
  {author} {\bibfnamefont {A.~D.}\ \bibnamefont {Mohite}}, \bibinfo {author}
  {\bibfnamefont {C.~C.}\ \bibnamefont {Galande}}, \bibinfo {author}
  {\bibfnamefont {P.~M.}\ \bibnamefont {Ajayan}}, \bibinfo {author}
  {\bibfnamefont {A.~M.}\ \bibnamefont {Dattelbaum}}, \bibinfo {author}
  {\bibfnamefont {H.}~\bibnamefont {Htoon}}, \bibinfo {author} {\bibfnamefont
  {A.~J.}\ \bibnamefont {Taylor}}, \ and\ \bibinfo {author} {\bibfnamefont
  {R.~P.}\ \bibnamefont {Prasankumar}},\ }in\ \href@noop {} {\emph {\bibinfo
  {booktitle} {{\itshape Nonlinear optics}}}}\ (\bibinfo  {publisher} {Optical
  Society of America},\ \bibinfo {year} {2011})\ p.\ \bibinfo {pages}
  {NFB2}\BibitemShut {NoStop}%
\bibitem [{\citenamefont {Bruus}\ and\ \citenamefont
  {Flensberg}(2004)}]{Bruus}%
  \BibitemOpen
  \bibfield  {author} {\bibinfo {author} {\bibfnamefont {H.}~\bibnamefont
  {Bruus}}\ and\ \bibinfo {author} {\bibfnamefont {K.}~\bibnamefont
  {Flensberg}},\ }\href@noop {} {\emph {\bibinfo {title} {{\itshape Many-body
  quantum theory in condensed matter physics: an introduction}}}}\ (\bibinfo
  {publisher} {Oxford University Press},\ \bibinfo {address} {Oxford},\
  \bibinfo {year} {2004})\BibitemShut {NoStop}%
\bibitem [{\citenamefont {Giuliani}\ and\ \citenamefont
  {Vignale}(2005)}]{Giuliani}%
  \BibitemOpen
  \bibfield  {author} {\bibinfo {author} {\bibfnamefont {G.~F.}\ \bibnamefont
  {Giuliani}}\ and\ \bibinfo {author} {\bibfnamefont {G.}~\bibnamefont
  {Vignale}},\ }\href@noop {} {\emph {\bibinfo {title} {{\itshape Quantum
  theory of the electron liquid}}}}\ (\bibinfo  {publisher} {Cambridge
  University Press},\ \bibinfo {address} {Cambridge},\ \bibinfo {year}
  {2005})\BibitemShut {NoStop}%
\bibitem [{\citenamefont {Melrose}(2008)}]{Melrose1Book}%
  \BibitemOpen
  \bibfield  {author} {\bibinfo {author} {\bibfnamefont {D.~B.}\ \bibnamefont
  {Melrose}},\ }\href@noop {} {\emph {\bibinfo {title} {{\itshape Quantum
  plasmadynamics: unmagnetized plasmas}}}},\ \bibinfo {series}
  {\textnormal{Lecture Notes in Physics}}, Vol.\ \bibinfo {volume} {735}\
  (\bibinfo  {publisher} {Springer},\ \bibinfo {address} {New York},\ \bibinfo
  {year} {2008})\BibitemShut {NoStop}%
\bibitem [{\citenamefont {Altland}\ and\ \citenamefont
  {Simons}(2010)}]{Altland}%
  \BibitemOpen
  \bibfield  {author} {\bibinfo {author} {\bibfnamefont {A.}~\bibnamefont
  {Altland}}\ and\ \bibinfo {author} {\bibfnamefont {B.}~\bibnamefont
  {Simons}},\ }\href@noop {} {\emph {\bibinfo {title} {{\itshape Condensed
  matter field theory}}}},\ \bibinfo {edition} {2nd}\ ed.\ (\bibinfo
  {publisher} {Cambridge University Press},\ \bibinfo {address} {Cambridge},\
  \bibinfo {year} {2010})\BibitemShut {NoStop}%
\bibitem [{\citenamefont {Adler}(1962)}]{Adler}%
  \BibitemOpen
  \bibfield  {author} {\bibinfo {author} {\bibfnamefont {S.~L.}\ \bibnamefont
  {Adler}},\ }\href@noop {} {\bibfield  {journal} {\bibinfo  {journal} {Phys.
  Rev.}\ }\textbf {\bibinfo {volume} {{\bfseries 126}}},\ \bibinfo {pages}
  {413} (\bibinfo {year} {1962})}\BibitemShut {NoStop}%
\bibitem [{\citenamefont {Nam}(1967)}]{Nam}%
  \BibitemOpen
  \bibfield  {author} {\bibinfo {author} {\bibfnamefont {S.~B.}\ \bibnamefont
  {Nam}},\ }\href {\doibase 10.1103/PhysRev.156.470} {\bibfield  {journal}
  {\bibinfo  {journal} {Phys. Rev.}\ }\textbf {\bibinfo {volume} {{\bfseries
  156}}},\ \bibinfo {pages} {470} (\bibinfo {year} {1967})}\BibitemShut
  {NoStop}%
\bibitem [{Note1()}]{Note1}%
  \BibitemOpen
  \bibinfo {note} {Strictly speaking, our derivation \protect \textup {\hbox
  {\mathsurround \z@ \protect \normalfont (\ignorespaces \ref
  {eq_Ohm_derived4}\unskip \@@italiccorr )}}--\protect \textup {\hbox
  {\mathsurround \z@ \protect \normalfont (\ignorespaces \ref
  {eq_Ohm_derived}\unskip \@@italiccorr )}} will run into difficulties if there
  are static magnetic fields present. These would be proportional to $\delta
  (\omega )$ in the frequency domain, and hence any formal manipulation with
  terms like $1/\omega $ would lead to singular results. Often, however, the
  influence of static magnetic fields on the conduction current is simply
  neglected. If this approximation is untenable, then the problem is not so
  much that Ohm's law looses its covariance. Instead, it looses its very
  validity giving way to a more general expansion of the induced current in
  terms of the external fields. We refer the interested reader to the
  discussion in \cite [Sec.~6.7]{ED1}.}\BibitemShut {Stop}%
\bibitem [{\citenamefont {Melrose}(1973)}]{Melrose73}%
  \BibitemOpen
  \bibfield  {author} {\bibinfo {author} {\bibfnamefont {D.~B.}\ \bibnamefont
  {Melrose}},\ }\href@noop {} {\bibfield  {journal} {\bibinfo  {journal}
  {Plasma Phys.}\ }\textbf {\bibinfo {volume} {{\bfseries 15}}},\ \bibinfo
  {pages} {99} (\bibinfo {year} {1973})}\BibitemShut {NoStop}%
\bibitem [{\citenamefont {Scharf}(1995)}]{ScharfQED}%
  \BibitemOpen
  \bibfield  {author} {\bibinfo {author} {\bibfnamefont {G.}~\bibnamefont
  {Scharf}},\ }\href@noop {} {\emph {\bibinfo {title} {{\itshape Finite quantum
  electrodynamics: the causal approach}}}},\ \bibinfo {edition} {2nd}\ ed.,\
  \textnormal{Texts and Monographs in Physics}\ (\bibinfo  {publisher}
  {Springer-Verlag},\ \bibinfo {address} {Berlin/Heidelberg},\ \bibinfo {year}
  {1995})\BibitemShut {NoStop}%
\bibitem [{\citenamefont {Scheck}(2005)}]{Scheck}%
  \BibitemOpen
  \bibfield  {author} {\bibinfo {author} {\bibfnamefont {F.}~\bibnamefont
  {Scheck}},\ }\href@noop {} {\emph {\bibinfo {title} {{\itshape Mechanics:
  from Newton's laws to deterministic chaos}}}},\ \bibinfo {edition} {4th}\
  ed.\ (\bibinfo  {publisher} {Springer},\ \bibinfo {address} {Berlin},\
  \bibinfo {year} {2005})\BibitemShut {NoStop}%
\bibitem [{\citenamefont {Kleinert}(2008)}]{Kleinert}%
  \BibitemOpen
  \bibfield  {author} {\bibinfo {author} {\bibfnamefont {H.}~\bibnamefont
  {Kleinert}},\ }\href@noop {} {\emph {\bibinfo {title} {{\itshape Multivalued
  fields in condensed matter, electromagnetism, and gravitation}}}}\ (\bibinfo
  {publisher} {World Scientific Publishing Co. Pte. Ltd.},\ \bibinfo {address}
  {Singapore},\ \bibinfo {year} {2008})\BibitemShut {NoStop}%
\bibitem [{\citenamefont {Streater}\ and\ \citenamefont
  {Wightman}(1964)}]{Streater}%
  \BibitemOpen
  \bibfield  {author} {\bibinfo {author} {\bibfnamefont {R.~F.}\ \bibnamefont
  {Streater}}\ and\ \bibinfo {author} {\bibfnamefont {A.~S.}\ \bibnamefont
  {Wightman}},\ }\href@noop {} {\emph {\bibinfo {title} {{\itshape PCT, spin
  and statistics, and all that}}}}\ (\bibinfo  {publisher} {W. A. Benjamin,
  Inc.},\ \bibinfo {address} {New York},\ \bibinfo {year} {1964})\BibitemShut
  {NoStop}%
\bibitem [{\citenamefont {Rebhan}(2012)}]{RebhanRel}%
  \BibitemOpen
  \bibfield  {author} {\bibinfo {author} {\bibfnamefont {E.}~\bibnamefont
  {Rebhan}},\ }\href@noop {} {\emph {\bibinfo {title} {{\itshape Theoretische
  Physik: Relativit\"atstheorie und Kosmologie}}}}\ (\bibinfo  {publisher}
  {Springer-Verlag},\ \bibinfo {address} {Berlin/Heidelberg},\ \bibinfo {year}
  {2012})\BibitemShut {NoStop}%
\bibitem [{\citenamefont {Jackson}(1999)}]{Jackson}%
  \BibitemOpen
  \bibfield  {author} {\bibinfo {author} {\bibfnamefont {J.~D.}\ \bibnamefont
  {Jackson}},\ }\href@noop {} {\emph {\bibinfo {title} {{\itshape Classical
  electrodynamics}}}},\ \bibinfo {edition} {3rd}\ ed.\ (\bibinfo  {publisher}
  {John Wiley \& Sons, Inc.},\ \bibinfo {address} {Hoboken, NJ},\ \bibinfo
  {year} {1999})\BibitemShut {NoStop}%
\bibitem [{\citenamefont {Tsang}(1997)}]{Tsang}%
  \BibitemOpen
  \bibfield  {author} {\bibinfo {author} {\bibfnamefont {T.}~\bibnamefont
  {Tsang}},\ }\href@noop {} {\emph {\bibinfo {title} {{\itshape Classical
  electrodynamics}}}}\ (\bibinfo  {publisher} {World Scientific Publishing Co.
  Pte. Ltd.},\ \bibinfo {address} {Singapore},\ \bibinfo {year}
  {1997})\BibitemShut {NoStop}%
\bibitem [{\citenamefont {Le~Bellac}\ and\ \citenamefont
  {L{\'e}vy-Leblond}(1973)}]{Bellac}%
  \BibitemOpen
  \bibfield  {author} {\bibinfo {author} {\bibfnamefont {M.}~\bibnamefont
  {Le~Bellac}}\ and\ \bibinfo {author} {\bibfnamefont {J.-M.}\ \bibnamefont
  {L{\'e}vy-Leblond}},\ }\href@noop {} {\bibfield  {journal} {\bibinfo
  {journal} {Il Nuovo Cimento}\ }\textbf {\bibinfo {volume} {{\bfseries 14
  B}}},\ \bibinfo {pages} {217} (\bibinfo {year} {1973})}\BibitemShut {NoStop}%
\bibitem [{\citenamefont {Rousseaux}(2013)}]{Rousseaux}%
  \BibitemOpen
  \bibfield  {author} {\bibinfo {author} {\bibfnamefont {G.}~\bibnamefont
  {Rousseaux}},\ }\href@noop {} {\bibfield  {journal} {\bibinfo  {journal}
  {Eur. Phys. J. Plus}\ }\textbf {\bibinfo {volume} {{\bfseries 128}}},\
  \bibinfo {pages} {1} (\bibinfo {year} {2013})}\BibitemShut {NoStop}%
\end{thebibliography}%

\end{document}